\documentclass[aps,prl,reprint,showpacs,floatfix,twocolumn]{revtex4}

\usepackage{amsmath,amsthm,amsfonts,amssymb,amscd}
\usepackage{t1enc}
\usepackage[latin1]{inputenc}
\usepackage[english]{babel}
\usepackage{babel}
\usepackage{graphics}
\usepackage{graphicx}
\usepackage[T1]{fontenc}
\usepackage{latexsym}

\usepackage{enumerate}

\newcounter{eqs}
\newenvironment{eqs}{\refstepcounter{eqs}\equation}{\tag{S\theeqs}\endequation} 
\newcounter{eqns}

\newcommand{\ket}[1]{\ensuremath{|{#1}\rangle}} 
\newcommand{\avg}[1]{\ensuremath{\langle{#1}\rangle}} 

\begin{document}

\title{Weak-value amplification as an optimal metrological protocol} 
\author{G. Bi\'e Alves, B. M. Escher, R. L. de Matos Filho, N. Zagury and L. Davidovich}
\affiliation{Instituto de F\'isica, Universidade Federal do Rio de Janeiro, P.O.Box 68528, Rio de Janeiro, RJ 21941-972, Brazil}
\begin{abstract}
The  implementation of weak-value amplification requires the pre- and post-selection of states of a quantum system, followed by the observation of the response of the meter, which interacts weakly with the system.  Data acquisition from the meter is conditioned to successful post-selection events. Here we derive  an optimal post-selection procedure for estimating the coupling constant between system and meter, and show that it leads both to weak-value amplification and to the saturation of the quantum Fisher information, under conditions fulfilled by all previously reported experiments on the amplification of weak signals. For most of the pre-selected states, full information on the coupling constant can be extracted from the meter data set alone, while for a small fraction of the space of pre-selected states,  it must be obtained from the post-selection statistics. \end{abstract}
\maketitle

\textit{Introduction}. The notion of weak-value amplification (WVA),  introduced in the pioneer work of Y. Aharonov, D. Z. Albert, and L. Vaidman \cite{spin100}, has been frequently associated with the possibility of amplifying weak signals, as small birefringence effects \cite{hulet,phsr}, the spin Hall effect of light \cite{spin-hall}, tiny deflections of light produced by moving mirrors in optical setups \cite{howell,howell2,howell-SNR,turner},  slow-velocity measurements \cite{viza}, small phase-shifts in interferometers \cite{howell-phase},  small time delays of light \cite{bruder}, tiny optical angular rotations \cite{boyd}, or the measurement of small frequency changes in the optical domain  \cite{howell-freq}. 
As shown in \cite{spin100} and \cite{rw}, WVA may also lead to exotic results. The  example in \cite{rw} has played a pioneer role in the development of the theory of quantum random walks \cite{rw2} .

The procedure for attaining an amplification of the weak value, which has as essential ingredient a conditional measurement procedure, can be divided into two steps: (i) the system to be measured, prepared in a pre-selected initial state, first interacts weakly with a meter, through a bilinear coupling -- quantified by a coupling constant g -- between observables $\hat{A}$ of the system and $\hat{M}$ of the meter, and then is post-selected in a predetermined state, usually taken as almost orthogonal to the initial state of the system; (ii) the weak value (real part or imaginary part) is determined by observation of the meter, whenever the post-selection in the predetermined state is successful.  In this procedure, the amplification of the weak-value is not deterministic. The interaction between system and meter is assumed to take place during a short time interval, so that the free evolution of  system plus meter can be safely neglected. 

The possibility of amplifying very weak signals  via WVA   leads quite naturally to the question as to whether  such measurements may be used to enhance metrological protocols that aim to estimate the coupling constant $g$. However,  such procedures may lead  not only to amplification of the signal, but also to the mitigation of the number of experimental data (statistics) that may be used to estimate $g$. This has led to debate on the possible advantages of weak measurements over the standard quantum-measurement procedure \cite{steinberg,combes,combes2,jordan,knee}. Proper treatment of this problem requires the machinery of quantum metrology, which establishes general bounds for the uncertainty in the estimation of parameters \cite{helstrom,iqF3,escher0,BJP}, defined by the mean-square estimation error, and expressed in terms of the corresponding quantum Fisher informations.   From \cite{iqF3,combes2} and the above discussion, it is clear that the amount of information on $g$ cannot be superior to that quantified by the corresponding quantum Fisher information. This is an upper bound valid for any kind of measurement, including weak measurements.   In spite of this, practical advantages of weak measurements have been pointed out \cite{steinberg,jordan}.

Here we address the formalism of WVA itself, and propose an optimized post-selection procedure, which actually saturates the quantum Fisher information corresponding to the estimation of $g$, in the weak-coupling limit, and can be applied to all previously reported experiments involving amplification of weak signals.    This procedure leads to a post-selected state  that  is not, in general,  quasi-orthogonal to the initial state, as opposed to the usual approach. We also show that proper handling of the conditions of weak coupling between quantum system  and meter  involves  a  limiting procedure concerning two small quantities, the coupling constant and the overlap between initial and post-selected states, which when tackled properly leads to results at variance with previously published analyses. These results imply that WVA, even though relying on a reduced data set, may lead, under proper choice of the detection procedure, to the same information on the parameter to be estimated as optimal quantum measurement protocols.  

\textit{Quantum metrological limits.} 
We look now for the ultimate precision limit in the estimation of $g$ for a system described by the weak-coupling Hamiltonian. We assume that the meter ${\cal M}$ and the quantum system ${\cal A}$ couple linearly through the interaction $\hat{H}_I(t)=\hbar g\delta(t-t_0) \hat{A}\hat{M}$,  $g$ taken to be positive and dimensionless, without loss of generality.  ${\cal A}$ and ${\cal M}$ are initially prepared in the state $|\Psi_i\rangle=|\psi_i\rangle\otimes|\phi_i\rangle$, where $|\psi_i\rangle$ is the initial quantum state of ${\cal A}$ and $|\phi_i\rangle$ is the initial state of ${\cal M}$.

If one estimates the value of a general parameter $x$ trough $\nu$ repeated measurements on the system that carries information about it, then the minimum reachable uncertainty on unbiased estimatives of the parameter is determined by the 
 Cram\'er-Rao limit \cite{MLE,fisher1,fisher2}: $\delta x \ge {1}/{\sqrt{\nu F(x)}}$. Here $\delta x=\langle \left(x-x_{\rm est}\right)^2\rangle^{1/2}$ is the mean-square estimation error, the average is taken over all possible experimental results, and $x_{\rm est}$ is an estimate of the parameter $x$, based on the observed data. $F(x)$ is the  \emph{Fisher information}, defined by
\begin{equation}\label{iF}
    F(x) = \sum_k \frac{1}{P_k(x)}\left[\frac{dP_k(x)}{dx}\right]^2,
\end{equation}
where $P_k(x)$ is the probability distribution of obtaining an experimental result $k$, assuming that the value of the parameter is $x$. The Fisher information $F(x)$ depends, through  $P_k(x)$, on the state of the system and on the measurement performed on it. 

The maximization of $F(x)$ over all possible measurements leads to the { \it quantum  Fisher information}~\cite{helstrom,iqF2,iqF3} $\mathcal{F}(x)$, which depends  only on the $x$-dependent state of the system, and yields the minimum possible value of $\delta x$. For a pure state $|\Psi(x)\rangle$, it is given by \cite{helstrom}
\begin{equation}\label{iqF3}
\mathcal{F}(x) = 4\left[\frac{d\langle\Psi(x)|}{dx}\frac{d|\Psi(x)\rangle}{dx} - \left|\frac{d\langle\Psi(x)|}{dx}|\Psi(x)\rangle\right|^2 \right]\,.
\end{equation}

For $\hat U(g)=\exp[-i\int \hat H_I(t)dt]=\exp(-ig\hat A\hat M)$, and $|\Psi_i\rangle=|\psi_i\rangle\otimes|\phi_i\rangle$,  the maximum amount of information on the coupling parameter $g$ that can be obtained by measurements on ${\cal A}+{\cal M}$ is then, according to \eqref{iqF3},
\begin{equation}
 \mathcal{F}(g)=4\left[ \langle {\hat A^2}\rangle \langle{\hat M^2}\rangle-\langle{\hat A}\rangle^2\langle{\hat M}\rangle^2 \right]\,,
\label{qFishersistemmeter}
\end{equation}
where from now on the averages of operators corresponding to ${\cal A}$ and ${\cal M}$ are taken respectively in the states $|\psi_i\rangle$ and $|\phi_i\rangle$. We compare now this expression to the one corresponding to the WVA protocol.

\textit{Parameter estimation with post-selection}. For the evolution corresponding to $\hat U(g)$, the probability of detecting system ${\cal A}$ in the state $|\psi_f\rangle$ immediately after $t_0$ is 
$p_f(g) = \|\langle\psi_f\vert \hat{U}(g) \vert\Psi_i\rangle\|^2$. If ${\cal A}$ is detected in the state $|\psi_f\rangle$,  ${\cal M}$ is left in the normalized state 
\begin{equation}\label{meter-final}
    |\phi_f(g)\rangle = \langle\psi_f|\hat{U}(g)|\psi_i\rangle|\phi_i\rangle/\sqrt{p_f(g)}.
\end{equation}
The original WVA strategy involves measuring the meter ${\cal M}$ only when the system ${\cal A}$ is post-selected in $|\psi_f\rangle$. The post-selection statistics -- described by the post-selection probability $p_ f(g)$ in the asymptotic limit $\nu\rightarrow\infty$ -- is ignored in the estimation of $g$. Full consideration of the post-selection procedure should take it into account. This can be described through a set of generalized measurement operators $\{ |\psi_f\rangle\langle\psi_f|\otimes \hat E_j , (\hat 1_{\cal A}-|\psi_f\rangle\langle\psi_f|)\otimes\hat 1_{\cal M}\}$, $j=1,2,\dots n$, where the set $\{\hat E_j\}$, with $\sum_{j=1}^n \hat E_j=\hat 1_{\cal M}$, acts on the states of $\cal M$ \cite{combes2,walmsley}. The corresponding Fisher information for the estimation of the coupling constant $g$, as defined by \eqref{iF},   is $F_{ps}(g)=F_m(g)+F_{p_f}(g)$,
where  
\begin{equation}\label{Fishermeter}
F_{m}(g) =p_f(g)\sum_{j=1}^{n} \frac{1}{P_j(g) }\left[ \frac{dP_j(g)}{dg}\right]^2\,,
\end{equation}
with $P_j(g) = \langle\phi_f(g)\vert\hat{E}_j\vert\phi_f(g)\rangle$, is the Fisher information associated to measurements on the state of the meter after post-selection, times the probability $p_f(g)$ that the post-selection succeeds, and
\begin{equation}
F_{p_f}(g)=\frac{1}{p_f(g)[1-p_f(g)]}\left[\frac{dp_f(g)}{dg}\right]^2 \label{Fisherproj}
\end{equation}
stands for the information on $g$ encoded in $p_f(g)$. 

The amount of information $F_m(g)$ quantifies the {\it performance} of the estimation for the original WVA. It takes into account both the enhancement provided by the post-selection, through the state \eqref{meter-final}, and the degradation due to the loss of statistical data, through the probability $p_f(g)$. $F_{p_f}(g)$, on the other hand, quantifies the amount of information on $g$ acquired from $p_f(g)$ itself. The total information $F_{ps}(g)$ is obtained with the best unbiased estimative of $g$ that considers all available data in the experiment, when the meter is monitored only if the post-selection of the system is successful.

 The maximal value ${\mathcal F}_m(g)$ of $F_m(g)$ over all POVM's acting in the Hilbert space of the meter is obtained by inserting $|\phi_f(g)\rangle$ into \eqref{iqF3}, with $x\equiv g$, and multiplying the result by $p_f(g)$, yielding
 \begin{equation}\label{{Fisherm}}
\mathcal{F}_{\rm m}(g)\!=\!
4\left[\langle\hat Q(g)^\dagger \hat Q (g)\rangle\! -\!\left|\langle \hat Q(g)^\dagger \hat O(g)\rangle\right |^2/p_f(g)\right],
\end{equation}
where $\hat O(g)\!=\!\langle\psi_f| e^{-ig\hat A \hat M}|\psi_i\rangle$ and $\hat Q(g)\!=\!\langle\psi_f |\hat A\hat M e^{-ig\hat A \hat M}|\psi_i\rangle$ are operators that act in the Hilbert space of the meter. 
Note that ${\mathcal F}_m(g)$ is a functional of $\vert\Psi_i\rangle$ and of $\vert\psi_f\rangle$, the post-selected state. We define ${\cal F}_{ps}(g)\equiv{\cal F}_m(g)+F_{p_f}(g)$. Since in all reported WVA experiments only the meter is measured, a challenging question is whether the quantum Fisher information given in \eqref {qFishersistemmeter} can be attained by  ${\mathcal F}_m(g)$ alone. If not, can this be accomplished by ${\cal F}_ {ps}(g)$?

In the next section we examine these questions, as we specialize these results to the weak-coupling regime. 

\textit{Weak-coupling regime with balanced meters.} We solve the problem of maximizing ${\mathcal F}_{ps}(g)$ over the state $\vert\psi_f\rangle$ in the weak-coupling limit, with the condition $\langle\hat M\rangle=0$ (balanced meter). Then $\Delta\equiv\langle\hat M^2\rangle^{1/2}$ is the standard deviation of the initial distribution of eigenvalues of $\hat M$, and the quantum Fisher information becomes ${\cal F}(g)=4\langle\hat A^2\rangle\Delta^2$.  Under these conditions, and for separable initial states, we shall show that it is always possible to find a post-selected state  $\vert\psi_f\rangle$, such that ${\cal F}_{ps}(g)$ reaches, up to first order in $g$, the  quantum Fisher information ${\cal F}(g)$. Those conditions, although restrictive, are fulfilled in all experiments aimed to amplify weak signals, reported so far \cite{hulet,phsr,spin-hall,howell,howell2,howell-SNR,turner,howell-freq}. We discuss first the situation where ${\cal F}_{m}(g)$ alone reaches  ${\cal F}(g)$.

For $g$ sufficiently small,  we show in the Supplemental Material \cite{sup} that  ${\mathcal F}_{m}(g)=4 \Delta^2|\langle\psi_i|\hat{A}\vert\psi_f\rangle|^2 [1+ O(g)]$. This implies that the ansatz  $\vert\psi_f^{\rm opt}\rangle=\hat{A}\vert\psi_i\rangle/\langle  \hat A^2\rangle^{1/2}$ leads to ${\mathcal F}_{m}(g)\rightarrow {\mathcal F}_{m}^{\rm opt}(g)= {\cal F}(g) + O(g^2)$, where the superscript ${\rm opt}$ specifies quantities corresponding to the above post-selection. Therefore, ${\mathcal F}_{m}^{\rm opt}(g)$ coincides with the quantum Fisher information, up to first order in $g$. For this optimal post-selection, the correction must be necessarily non-positive, independently of the sign of $g$, which excludes corrections of $O(g)$. 

Quantifying the meaning of ``$g$ sufficiently small''  requires  careful consideration of the relative magnitudes of $\delta\equiv \langle\psi_f^{\rm opt}|\psi_i\rangle=\langle\hat A\rangle/\langle\hat A^2\rangle^{1/2}$ and $g$, since $|\delta|$ may become much smaller than one, for some initial states, as discussed in the following. We shall show however  that, except for  very small values of $|\delta|$, as compared to $g$, the information from the meter is enough to saturate the quantum Fisher information ${\cal F}(g)$.

It is worthwhile to note that, for any  $\hat A$ and $|\psi_i\rangle$,  $\hat A|\psi_i\rangle=\langle\hat A\rangle|\psi_i\rangle+[\langle \hat A^2\rangle-\langle\hat A\rangle^2]^{1/2}|\psi_i\rangle_{\perp}$,  where   $|\psi_i\rangle_\perp$ is orthogonal to $|\psi_i\rangle$. Therefore, $|\psi_f^{\rm opt}\rangle$ is not, in general,  necessarily quasi-orthogonal to the initial state, which is typically assumed in the WVA literature to be the ideal post-selected state. Indeed, depending on $|\psi_i\rangle$, $|\psi_f^{\rm opt}\rangle$ may vary continuously from a state parallel to the initial state (when $|\psi_i\rangle$ is an eigenstate of $\hat A$) to a state orthogonal to $|\psi_i\rangle$ (when $\langle\hat A\rangle=0$).

The weak value of $\hat A$ is defined as  $A_w=\langle\psi_f\vert\hat A\vert\psi_i\rangle/ \langle\psi_f\vert\psi_i\rangle$ \cite{spin100}. For $|\psi_f\rangle=|\psi_f^{\rm opt}\rangle$, this becomes $A_w=\langle{\hat A^2}\rangle/\langle{\hat A}\rangle$, which, for any state that is not an eigenstate of $\hat A$,  is larger than the average value of $\hat A,$ i.e., there is amplification of the signal. In general, the magnitude of $A_w$ depends on the smallness of the absolute value of the scalar product of the pre- and post-selected states of the system.  
The limit of amplification, in order that  the weak value be well defined,  was discussed earlier in the literature \cite{sudarshan,kofman}. In particular, it is required that $g|A_w|\Delta\ll1$. Indeed, the signal obtained from the measurement of the meter is  de-amplified as one tries to get very close to the orthogonal post-selection, which has been named the {\it inverted region} \cite{kofman}. The   optimal post-selected state $|\psi_f^{\rm opt}\rangle$, which leads to the best precision in the estimation of $g$, does not provide the largest possible amplification established by  \cite{sudarshan}, since $|\delta|$ is not necessarily much smaller than one.   

We show now that, in two limiting cases, the information on $g$ gets concentrated either in ${\cal F}^{\rm opt}_m$ or $F^{\rm opt}_{p_f}$. We assume in the following balanced meters and the post selection of  $|\psi_f^{\rm opt}\rangle$. We have then \cite{sup}:

 {\bf (a)}  If $|\delta|\ll g\langle\hat A^2\rangle^{1/2}\Delta$, or equivalently $g|A_w|\Delta\gg1$, then $F_{p_f}^{\rm opt}(g)=\mathcal{F}(g)[1+O(\varepsilon)]$, and $\mathcal{F}_m^{\rm opt}(g)=O(\varepsilon)$, where $\varepsilon= \mbox{Max}\{|\delta|,g^2,(\delta/g)^2\}$. In this case, full information on $g$ can be obtained from the statistics of successful post-selection detections in $|\psi_f^{\rm opt}\rangle$. One should note, however, that this is the region of parameters for which the usual weak-value theory breaks down \cite{kofman}. Furthermore, this condition holds in a small region of overlaps $\delta$, since the convergence of the expansion in $g$ requires that $g\langle\hat A^2\rangle^{1/2}\Delta\ll 1$ \cite{kofman}. Typical experimental values of $g\langle\hat A^2\rangle^{1/2}\Delta$  range  from $10^{-3}$ \cite{spin-hall, viza}  to $10^{-8}$  \cite{howell}.

{\bf (b)} If  $|\delta|\gg g\langle\hat A^2\rangle^{1/2}\Delta$, or equivalently  $g|A_w|\Delta\ll1$ (regime of validity of the weak-value theory), 
$ \mathcal{F}_{m}^{\rm opt}(g)= \mathcal{F}(g)[1+\mathcal{O} (g^2)]$,  and $F_{p_f}^{\rm opt}(g)=\mathcal{O} (g^2)$.  Therefore, full information on $g$ is now obtained by considering just the best measurement on the pointer after post-selection. 
 
 One should note that condition {\bf (a)} includes the region of initial states where $|\psi_f^{\rm opt}\rangle$ becomes orthogonal to $|\psi_i\rangle$. This implies, surprisingly, that even though exact orthogonality is avoided in typical WVA treatments, it actually leads to saturation of the quantum Fisher information, with the information on $g$ fully concentrated in the statistics of successful post-selection events. Then, measurements on the meter,  the ones  considered in most WVA analysis,  yield no information on $g$.

All these results were obtained for post-selection in $|\psi_f\rangle=|\psi_f^{\rm opt}\rangle$. As shown in the Supplementary Material \cite{sup}, the choice $|\psi_f\rangle=|\psi_i\rangle$ also leads to ${\cal F}_{ps}(g)={\cal F}(g) + O(g^2)$. In this case, the information obtained from the statistics of successful post-selection becomes important in a broader region of initial states. One should note, however, that in this case $A_w=\langle\psi_i|\hat A|\psi_i\rangle$, and therefore there is no weak-value amplification. This shows that, even in a post-selection procedure,  WVA is not needed in order to increase the precision in the estimation of $g$. 

The example  discussed in the following illustrates the results of this section.

\textit{Weak-value amplification of a spin.} 
Consider a two-level system, which interacts with the meter in such a way that $\hat A =\hat \sigma_z$. We parametrize the pre- and post-selected states by  the angles $\{ \theta_i,\theta_f, \phi_i, \phi_f \}$, so that $|\psi_i\rangle= \cos(\theta_i/2) |0\rangle+  e^{i\phi_i} \sin(\theta_i/2)|1\rangle$    and $ |\psi_f\rangle= \cos(\theta_f/2)|0\rangle + e^{i\phi_f} \sin(\theta_f/2)|1\rangle$, where $|0\rangle$ and $|1\rangle$ are the eigenvectors of $\hat\sigma_z$ corresponding to the eigenvalues $+1$ and $-1$, respectively. Since $\hat A^2=\hat1$, the small-coupling condition requires that $g\Delta\ll 1$. 

According to the previous discussion,  the bound  $\mathcal{F}(g)$ can be approached when  $|\psi_f\rangle=\hat\sigma_z|\psi_i\rangle,$  $\langle \hat M \rangle=0$, and if terms of ${O}(g^2)$  are neglected.  For this post-selected state,   $\theta_f=\theta_i=\theta$ and $\phi\equiv\phi_f-\phi_i=\pi $ ($\delta=\cos\theta$). Notice that ${\cal F}_m^{\rm opt}(g)$, $F_{p_f}(g)$, and ${\cal F}_{ps}^{\rm opt}(g)$ are invariant under the transformation $\theta\rightarrow\pi-\theta$.

\begin{figure}[t]
\centering
\includegraphics[width=0.45\textwidth]{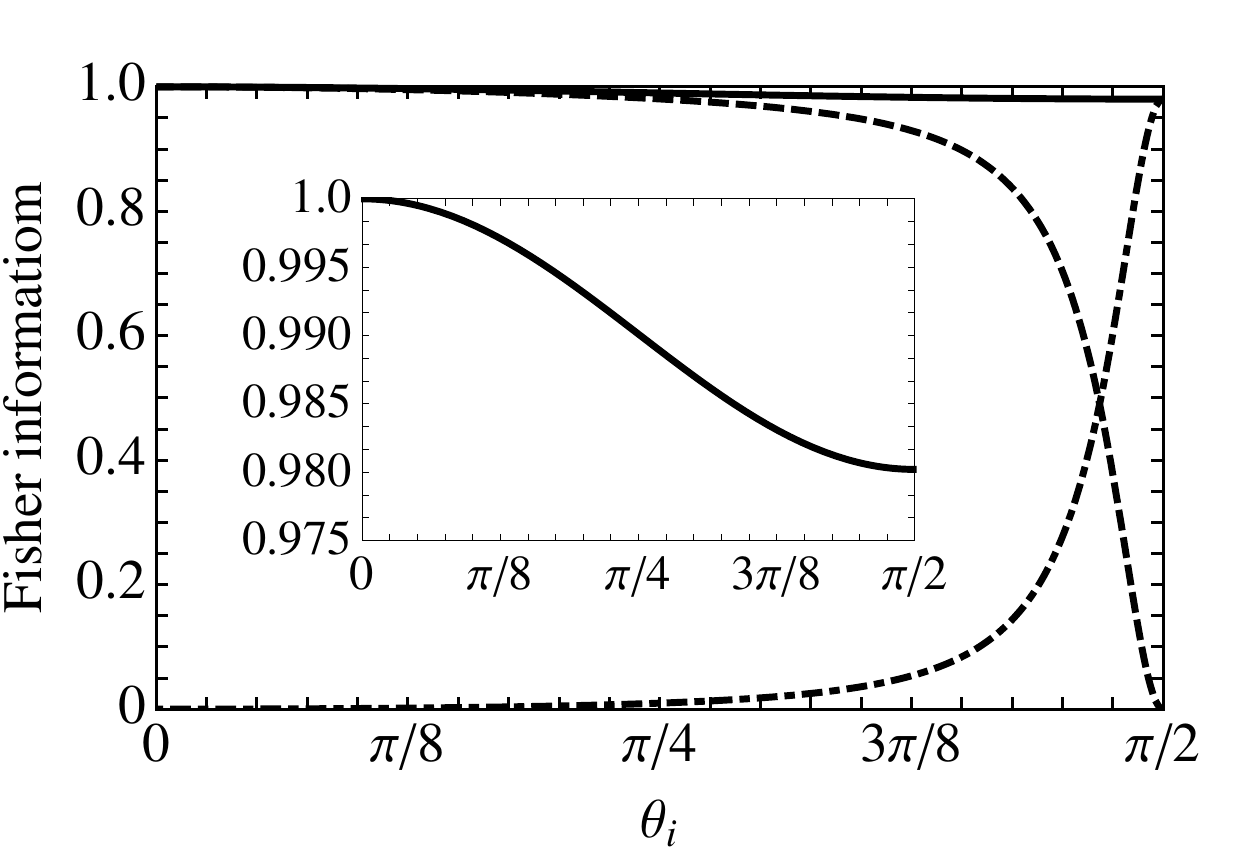}
\caption{Bounds for  the information from the meter,
$\mathcal{F}_{m}^{\rm opt}(g)$ (dashed line) and from the probability of post-selection, $F_{p_f}^{\rm opt}(g)$ (dotted-dashed line), normalized by the quantum Fisher information,  as a function of $\theta_i$, for $g\Delta=0.1$.  The full line, also displayed  in the inset, is the sum of the two contributions. It does not reach the value one, since saturation of ${\cal F}(g)$ is up to errors of $O(g^2)$.}
\label{fig1}
\end{figure}
  \begin{figure}[b]
\centering
\includegraphics[width=0.45\textwidth]{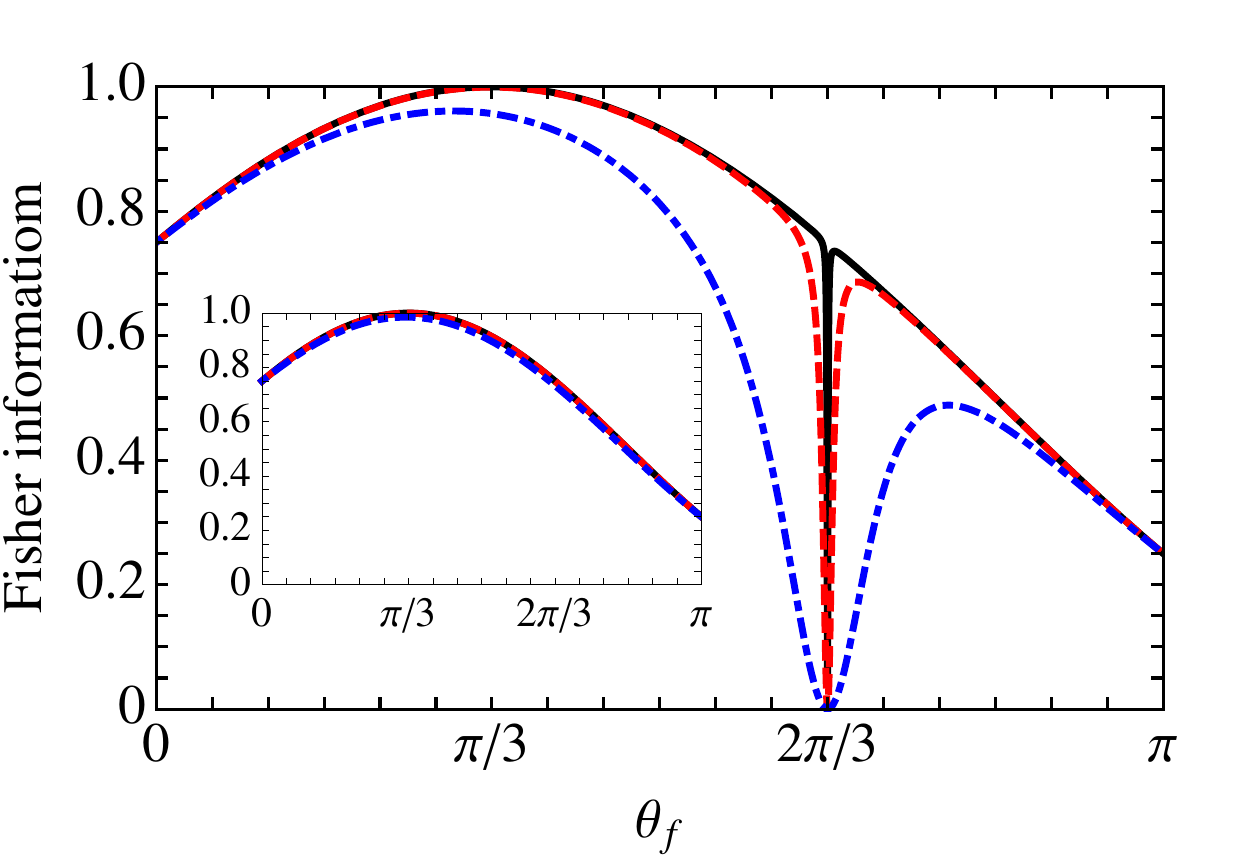}
\caption{ $\mathcal{F}_{m}(g)$  normalized to $\mathcal{F}(g)$   as a function of $\theta_f$ for $\theta_i=\pi/3$ and $\phi=\pi.$ The inset shows the full post-selected Fisher information ${\cal F}_{ps}(g)$,  showing that $F_{ p_f}(g)$  fills the dip of $\mathcal{F}_{m}(g)$.  $g\Delta=10^{-1}$ (dotted-dashed blue line), $10^{-2}$ (dashed red line), $10^{-3}$ (full black line).}
\label{fig2}
\end{figure}
Figure~\ref {fig1} illustrates the balance of information between meter and post-selection statistics by plotting the contributions of $\mathcal{F}_{m}^{\rm opt}(g)/{\cal F}(g)$ (dashed line) and $F_{p_f}^{\rm opt}(g)/{\cal F}(g)$ (dotted-dashed line) as a function of $\theta_i$, for $g\Delta=0.1$. We have assumed that the initial state of the meter is a pure state   with a Gaussian distribution of the eigenvalues of $\hat M$, with width $\Delta=\langle \hat M^2\rangle^{1/2}$.  For $|\delta|\ll g\Delta$ (case {\bf (a)}),   the major contribution to the quantum Fisher information comes from $F_{p_f}^{\rm opt}(g)$.  As $|\delta | $ increases,  the contribution  $\mathcal{F}_{m}^{\rm opt}(g)$  becomes more relevant (case {\bf (b)}). 
The point where the two contributions coincide is very close to $|\delta|=g\Delta$, or equivalently $|A_w|=(g\Delta)^{-1}$, as one should expect from the above analysis.

Figure \ref{fig2} illustrates the behavior of   $\mathcal{F}_{m}(g)/{\cal F}(g)$ and  ${\cal F}_{ps}(g)/{\cal F}(g)$ (inset), normalized to the quantum Fisher information, as a function of  $\theta_f$, for several values of $g\Delta$.  As discussed before,  $\theta_f=\theta_i$ corresponds to an optimal post-selection procedure, since then $|\psi_f^{\rm opt}\rangle=\hat\sigma_z|\psi_i\rangle$. As shown in Fig.~\ref{fig2}, there is a dip in $\mathcal{F}_{m}(g)$, which corresponds to the region where ${F}_{p_f}(g)$ must be taken into account.  This dip becomes wider as  $g\Delta$ increases, a behavior that is analytically described in the Supplemental Material \cite{sup}. The inset shows that the maximum of ${\cal F}_{ps}(g)$ is reached for $\theta_f=\theta_i\approx\pi/3$, for the values of $g\Delta$ considered.    For $g\Delta\ll 1$, the meter information dominates over  practically the whole range of values of $\theta_f$, except for a vary narrow dip, which is compensated by a corresponding sharp increase of the information in ${F}_{p_f}(g)$, so that the full post-selection Fisher information ${\cal F}_{ps}(g)$, displayed in the inset, has a smooth behavior. Post-selection in the initial state is discussed in \cite{sup}.

\textit{Conclusion.}  We have shown that the information on the coupling constant between system and meter, obtained through a post-selection procedure that leads to weak-value amplification, saturates the quantum Fisher information in the weak-coupling regime.   As the post- and pre-selected states get orthogonal to each other, the information on the parameter gets transferred from the meter to the post-selection statistics, which then plays the dominant role in the estimation protocol, albeit restricted to a small fraction of the space of pre-selected states. These results imply that one of the main advantages of weak-value amplification, namely the ability to amplify signals of the meter, does not spoil the estimation precision, even though it relies on the reduced data set corresponding to the conditional observation of the meter. 

\textit{Acknowledgments}. This research was supported by the Brazilian agencies FAPERJ, CNPq, CAPES, and the National Institute of Science and Technology for Quantum Information.

%
%
%


\appendix
\onecolumngrid

%

\begin{center}
{\huge Supplemental Material}
\end{center}

\section{Maximization of $F_m(g)$ and $F_{ps}(g)$ over all possible measurements on the meter.}

We assume here, as in the main text, that  system ${\cal A}$ and meter ${\cal M}$ are initially prepared in the separable state $|\Psi_i\rangle=|\psi_i\rangle\otimes|\phi_i\rangle$. The maximization of $F_{m}(g)$, and also of $F_{ps}(g)$, over all possible POVMs $\{\hat{E}_j\}$ yields the information ${\cal F}_{m}(g)$, and, respectively, ${\cal F}_{ps}(g)$. ${\cal F}_{m}(g)$ can be obtained through the expression of the quantum Fisher information of the state $\vert\phi(g)\rangle$ defined by Eq.~(5) of the main text. After a straightforward calculation, we get
\begin{eqs}\label{sfm1}
\mathcal{F}_m(g) = 4p_f(g)\left[\dfrac{d\langle\phi(g)\vert}{dg}\dfrac{d\vert\phi(g)\rangle}{dg} - \left\vert\dfrac{d\langle\phi(g)\vert}{dg}\vert\phi(g)\rangle\right\vert^2\right] = 4\left[\langle\hat{Q}^\dagger(g)\hat{Q}(g)\rangle-\dfrac{|\langle\hat{Q}^\dagger(g)\hat{O}(g)\rangle|^2}{p_f(g)}\right] \, ,
\end{eqs}
\begin{eqs}\label{sfp1}
F_{p_f}(g) = \dfrac{4\mbox{Im}^2\{\langle\hat{Q}^\dagger(g)\hat{O}(g)\rangle\}}{p_f(g)}+\dfrac{4\mbox{Im}^2\{\langle\hat{Q}^\dagger(g)\hat{O}(g)\rangle\}}{1-p_f(g)}\,,
\end{eqs}
where
\begin{eqs}
\hat{Q}(g)=\langle\psi_f\vert\hat{A}\hat{M}e^{-i g \hat{A}\hat{M}}\vert\psi_i\rangle,
\end{eqs}
\begin{eqs}
\hat{O}(g)=\langle\psi_f\vert e^{-i g \hat{A}\hat{M}}\vert\psi_i\rangle.
\end{eqs}
\begin{eqs} \label{spf1}
p_f(g)= \langle \hat O(g)^\dagger \hat O(g) \rangle\, ,
\end{eqs}
and we have used the notation $\langle \hat X \rangle $ to denote the average (over the initial state) in the Hilbert space where the operator $\hat X $ acts. The coupling constant $g$ is assumed to be dimensionless.

The above expressions imply that
\begin{eqs}\label{sfps}
\mathcal{F}_{ps}(g)={\cal F}_m(g)+F_{p_f}(g)=4\left[\langle\hat{Q}^{\dag}(g)\hat{Q}(g)\rangle-\dfrac{\mbox{Re}^2\{\langle\hat{Q}^{\dag}(g)\hat{O}(g)\rangle\}}{p_f(g)}+
  \dfrac{\mbox{Im}^2\{\langle\hat{Q}^{\dag}(g)\hat{O}(g)\rangle\}}{1-p_f(g)}\right]\, .
\end{eqs}

 Using the notation $(\hat{A}^n)_{fi} \equiv \langle\psi_f\vert\hat A^n\vert\psi_i\rangle$, assuming that $(\hat{A}^n)_{fi}$  is bounded, and defining $\delta\equiv\langle\psi_f \vert\psi_i\rangle $ as real and positive (without loss of generality), we have
\begin{eqs}
\begin{split}\label{pf}
&\langle  \hat{O}^\dagger(g)\hat{O}(g) \rangle=p_{f}(g)=\delta^{2}+ 2g \delta\mbox{Im}\{ \hat{A}_{fi} \} \langle\hat{M}\rangle
+g^{2}\left[|\hat A_{fi}|^{2}-
\delta \mbox{Re}\{(\hat A^2)_{fi}\} \right] \langle \hat{M^{2}} \rangle \\
&-g^3/3\, \mbox{Im}\{\delta (\hat{A}^3)_{fi} +3(\hat{A}^2)^{*}_{fi}\hat{A}_{fi} \} \avg{\hat{M}^3}
+g^4/12\left[ \mbox{Re}\{ \delta(\hat{A}^4 )_{fi} - 4\hat{A}_{fi}^*(\hat{A}^3)_{fi} \} + 3|(\hat{A}^2)_{fi}|^2 \right]\avg{\hat{M}^4} + {\cal O}(g^5)\, ,
\end{split}
\end{eqs}
\begin{eqs}
\begin{split}\label{scond1s}
&\langle  \hat{Q}^\dagger(g)\hat{Q}(g) \rangle = \vert\hat{A}_{fi}\vert^2\avg {\hat{M}^2} - 2g\mbox{Im}\{(\hat{A^2})_{fi}^*\hat{A}_{fi}\}\avg {\hat{M}^3}
+ g^2\left[   |(\hat{A^2})_{fi}|^2 - \mbox{Re}\{(\hat{A^3})_{fi}^*\hat{A}_{fi} \}\right] \avg {\hat{M}^4} \\
&-g^3/3\left[ \mbox{Im}\{(\hat A^4)_{fi}\hat A_{fi}^* - 3(\hat A^3)_{fi}(\hat A^2)_{fi}^*\} \right]\avg{\hat M^5} + g^4/12\left[\mbox{Re}\{(\hat A^5)_{fi}\hat A_{fi}^* - 4(\hat A^4)_{fi}(\hat A^2)_{fi}^*\} + 3|(\hat A^3)_{fi}|^2\right]\avg{\hat M^6} + {\cal O}(g^5)\, ,
\end{split}
\end{eqs}
and
\begin{eqs}\label{scond2s}
\begin{split}
&\langle  \hat{Q}^\dagger(g)\hat{O}(g) \rangle =\delta A_{fi}^*\langle\hat{M}\rangle+
 ig\left[\delta (\hat{A^2})_{fi}^* - |\hat{A}_{fi}|^2\right] \avg {\hat{M}^2}
+ g^2/2\left[2(\hat{A^2})_{fi}^*\hat{A}_{fi} - (\hat{A^2})_{fi}\hat{A}_{fi}^*
- (\hat{A^3})_{fi}^* \delta \right]\avg {\hat{M}^3} \\
&-ig^3/6\left[\delta(\hat A^4)_{fi}^* - 3(\hat A^3)_{fi}^*\hat A_{fi} + 3|(\hat A^2)_{fi}|^2 - \hat A_{fi}^*(\hat A^3)_{fi}\right]\avg{\hat M^4} \\
&+ g^4/24\left[\delta(\hat A^5)_{fi}^* - 4(\hat A^4)_{fi}^*\hat A_{fi} + 6(\hat A^3)_{fi}^*(\hat A^2)_{fi} - 4(\hat A^2)_{fi}^*(\hat A^3)_{fi} + \hat A_{fi}^*(A^4)_{fi}\right]\avg{\hat M^5} + {\cal O}(g^5)\, .
\end{split}
\end{eqs}
 We assume here that the above expansions converge, for $g$ sufficiently small. We will also assume, as in the main text, that the initial state of the meter satisfies the condition that we call {\it balanced meter}:
\begin{eqs}\label{meanM}
\avg {\hat{M}} = 0.
\end{eqs}

\section{Analysis of the Fisher informations ${\cal F}_m(g)$, and $F_{p_f}(g)$ as a function of the initial state of the system $\vert\psi_i\rangle$, for a general post-selected state $\vert\psi_f\rangle$}

In analyzing the behavior of (\ref{sfm1}, \ref{sfp1}, \ref{sfps}) we should consider  the dependence on $\delta$ of $p_f(g).$ It is easy to show that, using the balanced-meter condition, we may write
\begin{eqs} \label{pfexpansion}
p_{f}(g)=\delta^{2}+g^{2}Z(\delta)+R(g,\delta )\, ,
\end{eqs}
where $R(g,\delta)$ is of order $g^3$ and $Z(\delta)=\left[|\hat A_{fi}|^{2}-\delta {\rm Re}\{(\hat A^2)_{fi}\} \right] \langle \hat{M^{2}} \rangle$, which we assume to be different from zero. Thus,
\begin{eqs}
F_{p_{f}}(g) = \frac{4g^{2} Z(\delta)^{2} + {\cal O}(g^3)}{\delta^2 +g^{2}Z(\delta)+ {\cal O}(g^3)} + \frac{4g^{2}Z(\delta)^{2} + {\cal O}(g^3)}{1-\delta^2-g^{2}Z(\delta) + {\cal O}(g^3)}\, ,
\end{eqs}
\begin{eqs}\label{ifm2}
\begin{split}
{\cal F}_m(g) = 4\vert\hat{A}_{fi}\vert^2\avg {\hat{M}^2} - 8g\mbox{Im}\{(\hat{A^2})_{fi}^*\hat{A}_{fi}\}\avg {\hat{M}^3}+ 4g^2\left(   |(\hat{A^2})_{fi}|^2 - \mbox{Re}\{(\hat{A^3})_{fi}^*\hat{A}_{fi} \}\right) \avg {\hat{M}^4} + {\cal O}(g^3)\\ - \frac{4g^2\left[Z(\delta) ^2+ \delta^2 \text{Im}^2\{ (\hat{A^2})_{fi}\} \avg {\hat{M}^2}^2\right]+{\cal O}(g^3)}{\delta ^{2}+g^{2}Z(\delta)+{\cal O}(g^3)}\,.
\end{split}
\end{eqs}
 Comparison of  the magnitude of $\delta^2$ and $1-\delta^2$ with the two terms of $g^2Z(\delta)$ -- the largest terms in the denominators of ${\cal F}_m(g)$ and $F_{p_f}(g)$ -- leads to the following limiting cases:
\begin{enumerate}[i.]
    \item \label{caso_i} If $\delta^2 \gg g^2|\hat A_{fi}|^2\Delta^2$ (assuming $|\hat A_{fi}|\neq0$) and $1-\delta^2 \gg \mbox{max}\left\{g^2|\hat A_{fi}|^2\Delta^2, g^2|\mbox{Re}\{(\hat A^2)_{fi}\}|\Delta^2\right\}$, where $\Delta\equiv\langle\hat M^2\rangle^{1/2}$ is the standard deviation of the meter eigenvalues distribution, then $F_{p_f}\sim {\cal O}(g^2)$ and
        \begin{eqs}\label{sfm2}
        {\cal F}_m(g) = 4\vert\langle\psi_f\vert\hat{A}\vert\psi_i\rangle\vert^2\Delta^2+ {\cal O}(g).
        \end{eqs}

    \item \label{caso_ii} In the small region $\delta^2\lesssim g^2|\hat A_{fi}|^2\Delta^2$ ($|\hat A_{fi}|\neq0$) we should have $Z(\delta) \approx |\hat A_{fi}|^{2}\langle \hat{M^{2}}\rangle+\mathcal{O}(\delta)$ and $R(g,\delta)\approx \mathcal{O}(g^3)$. Hence $F_{p_f}(g)$ may be well approximated by
        \begin{eqs}\label{Fo}
        F_{p_f}(g) = \frac{4g^{2} |\hat A_{fi} |^{4}\Delta^4 }{\delta^2 +g^{2}|\hat{A}_{fi}
        |^{2}\Delta^2}+{\cal O}(g)\, ,
        \end{eqs}
        which  has  the maximum value $ 4|\hat A_{fi} |^{2}\Delta^2$  at $\delta=0$ (post-selected state orthogonal to the initial state).
        Analogously, we may neglect the term $\delta^2 \text{Im}^2\{ (\hat{A^2})_{fi}\} \Delta^4$ in \eqref{ifm2} which will contribute in order $g^4$ (at most) in the numerator, and write
        \begin{eqs}\label{sfm}
        {\cal F}_m(g) = 4 \frac{\delta^2|\hat{A}_{fi}|^2 \Delta^2 }{\delta ^{2}+g^{2}|\hat{A}_{fi}|^{2}\Delta^2} +  \mathcal{O}(g).
        \end{eqs}
        Equations \eqref{Fo} and \eqref{sfm}  explain the sharp dip in ${\cal F}_m$ captured in Fig.~2 of the main text, around the point where $\ket{\psi_f}$ gets orthogonal to $\ket{\psi_i}$,  and show that in the same region $F_{p_f}$ has a sharp peak, so that their sum recovers a smooth function, as shown in the same figure.

\item The situation when $1-\delta^2 \lesssim \mbox{max}\left\{g^2|\hat A_{fi}|^2\Delta^2, g^2|\mbox{Re}\{(\hat A^2)_{fi}\}|\Delta^2\right\}$ is more subtle. We analyze it in the following, for specific choices of the post-selected state.

\end{enumerate}

\section{Analysis of the Fisher informations ${\cal F}_{ps}(g)$, ${\cal F}_m(g)$, and $F_{p_f}(g)$ as a function of the initial state of the system $\vert\psi_i\rangle$, for the post-selected state $\vert\psi_f\rangle=\vert\psi_f^{\rm opt}\rangle$.}

We consider now that the post-selected state of the system ${\cal A}$ is chosen as $\vert\psi_f\rangle=\vert\psi_f^{\rm opt}\rangle=\hat A|\psi_i\rangle/\langle\hat A^2\rangle^{1/2}$. As before, we assume the initial state of the meter to be such that $\langle\hat M\rangle=0$.

We focus on the physically relevant quantities, which are the information extracted from the meter, $\mathcal{F}_m(g)$, given by \eqref{sfm1} and from the post-selection probability, $F_{p_f}(g)$, as expressed by \eqref{sfp1}. Assuming the convergence of the expansions around $g=0$, one has
\begin{eqs}\label{QQ}
\langle\hat{Q}^{\dag}(g)\hat{Q}(g)\rangle = \langle\hat{A}^2\rangle\langle\hat{M}^2\rangle  - g^2\left[1 - \dfrac{\langle\hat{A}^3\rangle^2}{\langle\hat{A}^2\rangle\langle\hat{A}^4\rangle}\right]\langle\hat{M}^4\rangle\langle\hat{A}^4\rangle + {\cal O}(g^4)\,,
\end{eqs}
\begin{eqs}\label{sseconds}
 \dfrac{\mbox{Re}^2\{\langle\hat{Q}^{\dag}(g)\hat{O}(g)\rangle\}}{p_f(g)} = \dfrac{\dfrac{1}{4}\left(1-\delta\dfrac{\langle\hat{A}^4\rangle}{\langle\hat{A}^2\rangle^{1/2}\langle\hat{A}^3\rangle}\right)^2\langle\hat{A}^3\rangle^2\langle\hat{M}^3\rangle^2 g^4 +{\cal O}(g^6)}{\delta^2+\left(1-\delta\dfrac{\langle\hat{A}^3\rangle}{\langle\hat{A}^2\rangle^{3/2}}\right)\langle\hat{A}^2\rangle\langle\hat{M}^2\rangle g^2+{\cal O}(g^4)}\,,
\end{eqs}
\begin{eqs}\label{sthird}
 \dfrac{\mbox{Im}^2\{\langle\hat{Q}^{\dag}(g)\hat{O}(g)\rangle\}}{p_f(g)} = \dfrac{\left[1-\delta\dfrac{\langle\hat{A}^3\rangle}{\langle\hat{A}^{2}\rangle^{3/2}}\right]^2\langle\hat{M}^2\rangle^2\langle\hat{A}^2\rangle^2 g^2 + {\cal O}(g^4)}{\delta^2+\left(1-\delta\dfrac{\langle\hat{A}^3\rangle}{\langle\hat{A}^2\rangle^{3/2}}\right)\langle\hat{A}^2\rangle\langle\hat{M}^2\rangle g^2+{\cal O}(g^4)}\,,
\end{eqs}
\begin{eqs}\label{stermoi}
 \dfrac{\mbox{Im}^2\{\langle\hat{Q}^{\dag}(g)\hat{O}(g)\rangle\}}{1-p_f(g)} = \dfrac{\left[1-\delta\dfrac{\langle\hat{A}^3\rangle}{\langle\hat{A}^2\rangle^{3/2}}\right]^2\langle\hat{M}^2\rangle^2\langle\hat{A}^2\rangle^2 g^2 + {\cal O}(g^4)}{1 - \delta^2 - \left(1-\delta\dfrac{\langle\hat{A}^3\rangle}{\langle\hat{A}^2\rangle^{3/2}}\right)\langle\hat{A}^2\rangle\langle\hat{M}^2\rangle g^2+{\cal O}(g^4)}\,.
\end{eqs}
where $\delta\equiv\langle\psi_f^{\rm opt}|\psi_i\rangle= \langle\hat A\rangle/\langle\hat A^2\rangle^{1/2}$.

We show now that ${\cal F}_{ps}(g)$ saturates the quantum Fisher information, up to terms of first order in $g$. We notice from \eqref{QQ} that the first term on the right-hand side of \eqref{sfps} already saturates the quantum Fisher information ${\cal F}$, up to first order in $g$. From \eqref{sseconds}, the second term on the right-hand side of \eqref{sfps} is at most of ${\cal O}(g^2)$. Furthermore, the third term on the right-hand side of \eqref{sfps} is always positive, and therefore must be of  ${\cal O}(g^2)$, since ${\cal F}_{ps}$  cannot be larger than ${\cal F}$. Therefore,
\begin{eqs}
{\cal F}_{ps}(g) = 4 \langle\hat{A}^2\rangle\Delta^2 + {\cal O}(g^2)\,.
\end{eqs}

We analyze now two limiting cases, for which the information on $g$ is obtained  either from the meter or from the post-selection statistics.

{\bf (a)} For  $|\delta|=\langle \hat A\rangle/\langle\hat A^2\rangle^{1/2} \ll g\langle\hat A^2\rangle^{1/2}\Delta$,  the contribution from  \eqref{sseconds} is of ${\cal O}(g^2)$, as well as that from \eqref{stermoi}. On the other hand, \eqref{sthird}  contributes with ${\cal O}(g^0)$ in this limit.   In this regime, and assuming also that $\delta\langle\hat A^3\rangle/\langle\hat A^2\rangle^{3/2}\ll 1$,  \eqref{sthird} can be written as
\begin{eqs}\label{simb}
 \dfrac{\mbox{Im}^2\{\langle\hat{Q}^{\dag}(g)\hat{O}(g)\rangle\}}{p_f(g)} = \dfrac{ g^2\langle\hat{A}^2\rangle^2\Delta^4}{\delta^2 +  g^2 \langle\hat{A}^2\rangle\Delta^2} + {\cal O}\left(\mbox{Max}\{\delta, g^2\}\right) \simeq \avg{\hat A^2}\Delta^2 + {\cal O}\left[\mbox{Max}\{\delta,g^2,(\delta/g)^2\}\right]\,,
\end{eqs}such that we end up with
\begin{eqs}\label{iFm-a}
    \mathcal{F}_m(g) = \mathcal{O}\left[\mbox{Max}\{\delta,g^2,(\delta/g)^2\}\right] \,,
\end{eqs}
\begin{eqs}\label{iFpf-a}
    F_{p_f}(g) = 4\avg{\hat{A}^2}\Delta^2 + {\cal O}\left[\mbox{Max}\{\delta,g^2,(\delta/g)^2\}\right]\,.
\end{eqs}
This expression coincides, up to first order in $g$, with the quantum Fisher information. Therefore, in this limit, the information on $g$ is obtained solely from the post-selection statistics.

{\bf (b)} For $|\delta| \gg g\avg{\hat{A}^2}^{1/2}\Delta$, the contribution from \eqref{sseconds} is of ${\cal O}(g^4)$ and that of \eqref{sthird} is of ${\cal O}(g^2)$.

We show now that the contribution from \eqref{stermoi} is of ${\cal O}(g^2)$.  This is trivially true if  $1-\delta^2$ is not much smaller than one: then, it will be the dominating term in the denominator, and the right-hand side of \eqref{stermoi} will be of ${\cal O}(g^2)$. We show in the following that this still holds if $1-\delta^2\ll1$. In the limit $\delta\to1$,  $\ket{\psi_i}\to\ket{a}$, where $\ket{a}$ is some eigenstate of $\hat{A}$ with eigenvalue $a$. Therefore, for small values of $1-\delta^2$, $\ket{\psi_i}$ should be of the form
\begin{eqs}\label{estado-inicial}
    \ket{\psi_i^\epsilon}=\frac{\ket{a}+\epsilon\ket{b}}{\sqrt{1+\epsilon^2}}\,,
\end{eqs}
where $\hat A |a \rangle=a |a\rangle, $ $\langle b|a\rangle=0$ and $\epsilon$ is chosen as real (without loss of generality), with $\epsilon \ll 1$.   This implies that $\delta^2=\langle\psi_i^\epsilon|\hat A|\psi_i^\epsilon\rangle^2/\langle\psi_i^\epsilon|\hat A^2|\psi_i^\epsilon\rangle=1-{\cal O}(\epsilon^2)$, that is, $1-\delta^2={\cal O}(\epsilon^2)$.
Also, the term of ${\cal O}(g^2)$ in the denominator of \eqref{stermoi} can be written as
\begin{eqs} g^2Z(\delta)\equiv\left(1-\delta{\langle\hat{A}^3\rangle}/{\langle\hat{A}^2\rangle^{3/2}}\right)\langle\hat{A}^2\rangle\langle\hat{M}^2\rangle g^2 = \frac{g^2\epsilon^2 \langle  \hat M^2  \rangle \left[2a^2 ( \hat A^2)_{bb} -a ( \hat A^3 )_{bb} -a^3  \hat A _{bb} +{\cal O}(\epsilon^2)\right]}{(1+ \epsilon^2)(a^2+\epsilon^2 ( \hat A^2 )_{bb})}  ={\cal O}(\epsilon^2 g^2)\,,\end{eqs}
so that
\begin{eqs}\label{eqsepsilon}
\frac{g^2|Z(\delta)|}{1-\delta^2}={\cal O}(g^2),
\end{eqs}
implying that, in the denominator of \eqref{stermoi}, the term $1-\delta^2$ always dominates over $g^2Z(\delta)$. Furthermore, in the same region $1-\delta^2\ll 1$, the numerator of \eqref{stermoi} is of ${\cal O}(\epsilon^4 g^2)$, so that \eqref{stermoi} is indeed
 of ${\cal O}(g^2)$. From \eqref{sfm1} and \eqref{sfp1}, one has then, for $|\delta| \gg g\avg{\hat{A}^2}^{1/2}\Delta$,
\begin{eqs}\label{iFm-b}
    \mathcal{F}_m(g) = 4\avg{\hat{A}^2}\Delta^2 + \mathcal{O}(g^2)\,,
\end{eqs}
and
\begin{eqs}\label{iFpf-b}
    F_{p_f}(g) = \mathcal{O}(g^2)\,.
\end{eqs}
Therefore, in this limit,  the information on $g$ stems from the meter alone.

\section{Analysis of the Fisher informations ${\cal F}_{ps}(g)$, ${\cal F}_m(g)$, and $F_{p_f}(g)$ as a function of the initial state of the system $\vert\psi_i\rangle$, for the post-selected state $\vert\psi_f\rangle=\vert\psi_i\rangle$.}

The Fisher informations ${\cal F}_{ps}(g)$, ${\cal F}_m(g)$, and $F_{p_f}(g)$ can be analyzed for $\vert\psi_f\rangle=\vert\psi_i\rangle$, in the limit of weak coupling, for $\langle\hat{M}\rangle=0$, by using the expansions \eqref{pf}, \eqref{scond1s}, and \eqref{scond2s}). For $\vert\psi_f\rangle=\vert\psi_i\rangle$, these expansions yield, respectively,
\begin{eqs}
\langle\hat{Q}^\dagger(g)\hat{Q}(g)\rangle = \langle\hat{A}\rangle^2\langle\hat{M}^2\rangle + g^2\left(\langle\hat{A}^2\rangle^2 - \langle\hat{A}^3\rangle\langle\hat{A}\rangle\right)\langle\hat{M}^4\rangle + {\cal O}(g^4),
\end{eqs}
\begin{eqs}
\begin{split}
\langle\hat{Q}^\dagger(g)\hat{O}(g)\rangle = ig\left(\langle\hat{A}^2\rangle - \langle\hat{A}\rangle^2\right) \langle\hat{M}^2\rangle + \dfrac{g^2}{2}\left( \avg{\hat{A}^2}\avg{\hat{A}} - \avg{\hat{A}^3}\right)\langle\hat{M}^3\rangle +  {\cal O}(g^3),
\end{split}
\end{eqs}
\begin{eqs}
p_f(g) = 1 - g^2\left( \langle\hat{A}^2\rangle - \langle\hat{A}\rangle^2 \right)\langle\hat{M}^2\rangle +  {\cal O}(g^4).
\end{eqs}

One gets, up to order $g^2$:
\begin{eqs}\label{iFm2}
\mathcal{F}_m(g) =4\avg{\hat{A}}^2\avg{\hat{M}^2} + g^2\avg{\hat{M}^4}\avg{\hat{A}^2}^2\left(1-\frac{\avg{\hat{A}}\avg{\hat{A}^3}}{\avg{\hat{A}^2}^2}\right) - \frac{4\left(1-\frac{\avg{\hat{A}}^2}{\avg{\hat{A}^2}}\right)^2\avg{\hat{A}^2}^2\avg{\hat{M}^2}^2g^2 + {\cal O}(g^4)}{1+\left(\frac{\avg{\hat{A}}^2}{\avg{\hat{A}^2}}-1\right)\avg{\hat{A}^2}\avg{\hat{M}^2}g^2 + {\cal O}(g^4)}+{\cal O}(g^4)\,,
\end{eqs}
\begin{eqs}\label{iF-pf-2}
F_{p_f}(g) = \frac{4g^2\avg{\hat{M}^2}^2\avg{\hat{A}^2}^2\left(1-\frac{\avg{\hat{A}}^2}{\avg{\hat{A}^2}}\right)^2+{\cal O}(g^4)}{g^2\avg{\hat{M}^2}\avg{\hat{A}^2}\left(1-\frac{\avg{\hat{A}}^2}{\avg{\hat{A}^2}}\right)+{\cal O}(g^4)} + {\cal O}(g^2)\,.
\end{eqs}

Therefore,
\begin{eqs}
{\cal F}_m(g) = 4 \langle\hat{A}\rangle^2\Delta^2 + {\cal O}(g^2),
\end{eqs}
and
\begin{eqs}
F_{p_f}(g) = 4 \left( \langle\hat{A}^2\rangle - \langle\hat{A}\rangle^2 \right)\Delta^2 + {\cal O}(g^2),
\end{eqs}
so that the total quantum Fisher information corresponding to the post-selection strategy is
\begin{eqs}
{\cal F}_{ps}(g) = 4 \langle\hat{A}^2\rangle\Delta^2 + \mathcal{O}(g^2),
\end{eqs}
which shows that the post-selection in the state $|\psi_f\rangle=|\psi_i\rangle$ also saturates the quantum Fisher information, up to terms of first order in $g$. One should note, however, that in this case  the repartition of information between the meter and the post-selection statistics differs markedly from that corresponding to the post-selection strategy previously discussed:  in particular, even though the probability of post-selection is close to one, the information on $g$ is obtained from the post-selection statistics alone if the initial state of the system is such that $\langle\hat A\rangle=0$.

\section{Exact expressions for the Fisher information ${\cal F}_{ps}(g)$, ${\cal F}_m(g)$, and $F_{p_f}(g)$ as functions of the initial state $\vert\psi_i\rangle$ of a two-level quantum system (qubit).}

We take $U(g)=e^{-ig\hat\sigma_3 \hat M}$, where $\hat\sigma_3$ is a Pauli operator of the two-level system and $\hat M$ is an operator  of the meter. We assume a  balanced meter with an initial  Gaussian probability distribution of eigenvalues of $\hat M$, with a width given by $\Delta=\sqrt{\langle \hat M^2 \rangle}$. The initial and post-selected states are parametrized as  $\ket{\psi_i}=\cos(\theta_i/2)\ket{0} +\exp(i\phi_i)\sin(\theta_i/2)\ket{1}$,  $\ket{\psi_f}=\cos(\theta_f/2)\ket{0} + e^{i\phi_f}\sin(\theta_f/2)\ket{1}$, where $|0\rangle$ and $|1\rangle$ are eigenvectors of $\hat\sigma_3$ corresponding respectively to the eigenvalues $+1$ and $-1$.  We may easily obtain analytic expressions for the quantities in \eqref{sfm1}, \eqref{sfp1}, and \eqref{spf1}. Defining $A=\cos(\theta_i/2)\cos(\theta_f/2)$ and $B=e^{i\phi}\sin(\theta_i/2)\sin(\theta_f/2)$, where $\phi\equiv\phi_f-\phi_i$, one has:
 \begin{eqs}\label{pfqubit}
 p_f(g)=A^2+|B|^2+2 A\text{Re}\{B\}e^{-2g^2 \Delta^2}\,,
 \end{eqs}
 \begin{eqs}\label{Fmqubit}
 \begin{split}
  \mathcal {F}_m (g)&=4\Delta^2 \Bigg\{ A^2+|B|^2-2A\text{Re}\{B\}e^{-2g^2 \Delta^2} (1-4g^2\Delta^2)\\
  -&\frac{16 A^2\text{Re}^2\{B\} g^2 \Delta^2    e^{-4\Delta^2 g^2} }{ A^2+|B|^2 + 2A\text{Re}\{B\}e^{-2g^2 \Delta^2} } \Bigg\}\,,
 \end{split}
 \end{eqs}
  and
 \begin{eqs}\label{Fpqubit}
 \begin{split}
 F_{p_f}(g)     &=\frac{64 g^2\Delta^4 A^2 \text{Re}^2\{B\} e^{-4g^2 \Delta^2}}{p_f(g)(1-p_f(g)) }\,.
 \end{split}
 \end{eqs}
\subsection{Analytical results for the post-selection state $\ket{\psi_f}=\vert\psi_f^{\rm opt}\rangle$}
In this case, one should take $\theta_i=\theta_f=\theta$ and $\phi=\pi,$ so that $A= \cos^2(\theta/2), B=-\sin^2(\theta/2).$ From \eqref{Fmqubit} and \eqref{Fpqubit},  one gets
\begin{eqs}\label{iFm-qubit}
\frac{\mathcal{F}_m(g)}{\mathcal{F}} = \frac{1}{2}(1+\cos^2\theta) + \frac{1}{2}(1-4g^2\Delta^2)e^{-2g^2\Delta^2}\sin^2\theta - \frac{2g^2\Delta^2 e^{-4g^2\Delta^2}\sin^4\theta}{1+\cos^2\theta-e^{-2g^2\Delta^2}\sin^2\theta}\,,
\end{eqs}
\begin{eqs}\label{iF-pf-qubit}
\frac{F_{p_f}(g)}{\mathcal{F}} = \frac{4g^2\Delta^2 e^{-4g^2\Delta^2}\sin^2\theta}{(1+\cos^2\theta-e^{-2g^2\Delta^2}\sin^2\theta)(1+e^{-2g^2\Delta^2})}\, .
\end{eqs}
 For $g\Delta \ll1$, expand \eqref{iFm-qubit} and \eqref{iF-pf-qubit} and obtain
\begin{eqs}\label{check-meter}
    \frac{\mathcal{F}_m(g)}{\mathcal{F}}
    =\frac{\delta^2}{\delta^2+(1-\delta^2)g^2\Delta^2} \left[1+{\cal O}(g^2)\right]\,,
\end{eqs}
\begin{eqs}\label{check-count}
    \frac{F_{p_f}(g)}{\mathcal{F}}=\frac{(1-\delta^2)g^2\Delta^2}{\delta^2+(1-2\delta^2)g^2\Delta^2}\left[1+{\cal O}(g^2)\right] \,,
\end{eqs}
where  $\delta^2=\cos^2\theta$. We notice that $F_{p_f}$  contributes to the Fisher information  only  in a very small region $\Delta\theta \approx \mathcal{O}(g\Delta)$ near the equatorial plane in the
Bloch sphere. Outside this region $\mathcal{F}_m(g)/\mathcal{F}(g)\approx1.$  Indeed $\mathcal{F}_m(g)$ is larger than $F_{p_f}(g)$ as $\theta$ increases from zero up to  the  value $(\pi/2- g\Delta)$, when their contributions  to the Fisher information coincide, if one neglects contributions of ${\cal O}(g^2)$.

\subsection{Analytical results for the post-selection state \ket{\psi_f}=\ket{\psi_i}}

In this case, one should take $\theta_i=\theta_f=\theta$ and $\phi=0.$ From \eqref{Fmqubit} and \eqref{Fpqubit}  one gets
\begin{eqs}\label{iFm-qubit2}
\frac{\mathcal{F}_m(g)}{\mathcal{F}} = \frac{1}{2}(1+\cos^2\theta)
-\frac{1}{2}(1-4g^2\Delta^2)e^{-2g^2\Delta^2}\sin^2\theta - \frac{2g^2\Delta^2 e^{-4g^2\Delta^2 }\sin^4\theta}{1+\cos^2\theta+e^{-2g^2\Delta^2 }\sin^2\theta}\,,
\end{eqs}
\begin{eqs}\label{iF-pf-qubit2}
\frac{F_{p_f}(g)}{\mathcal{F}} = \frac{4g^2\Delta^2 e^{-4g^2\Delta^2}\sin^2\theta}{(1+\cos^2\theta+e^{-2g^2\Delta^2}\sin^2\theta)(1-e^{-2g^2\Delta^2})}\,.
\end{eqs}
Expansion of these expressions for $g\Delta\ll 1$ leads to
\begin{eqs}
    \frac{\mathcal{F}_m(g)}{\mathcal{F}} \approx  \cos^2\theta -g^2\Delta^2( \sin^4\theta- 3 \sin^2\theta)\,,
\end{eqs}
\begin{eqs}
    \frac{F_{p_f}(g)}{\mathcal{F}} \approx \sin^2\theta+g^2\Delta^2(\sin^4\theta-3\sin^2\theta) )\,.
\end{eqs}
Therefore, in this case the Fisher information corresponding to the post-selection statistics must be taken into account over a broader range of initial states, as compared to the post-selection strategy considered before.


\end{document}